# Interactions Between Evanescent Photons and Environment

Moses Fayngold

*Department of Physics, New Jersey Institute of Technology, Newark, NJ 07102*

The evanescent waves named as EW1, EW2, EW3 are described in 3 respective experimental setups: 1) total internal reflection; 2) scattering on an inhomogeneous planar target; and 3) propagation along a waveguide. Some interactions are considered between EW2 and the environment. The latter may include a beam of probing particles and/or the screen on which the EW2 are formed. Some new properties of EW are described, such as complex energy eigenvalues in case of a movable screen, and evanescence exchange between the interacting objects. This reveals the connection between evanescent states and the Gamow states of the studied system. The 4-momentum exchange between EW2 and the probe is highly selective and may collapse the superposition of studied EW2- eigenstates to a single EW-eigenstate of the probing particle. Possible imprints of EW2 in the far field are briefly discussed and a simple experiment is suggested for their observation.

.

# 1. Introduction

The simplest evanescent states (ES) are momentum eigenstates of a particle within some classically inaccessible (for the given energy and momentum) region of space with a constant potential. Their distinction from a regular momentum eigenstate (de Broglie's wave) is a complex vector eigenvalue and the existence of at least one border truncating the free-space region, with the corresponding boundary conditions [1-9]. In this work, we assume such particle to be primarily a photon, albeit its evanescence can be transferred in the process of interaction to other objects.

A well-studied case of evanescent waves (EW) appears in the total internal reflection (TIR). This is the simplest type – for a plane interface between two media and for an appropriate de Broglie's wave in the input, it produces only one EW eigenstate and only on the transmission side in the output. Therefore it was labeled as EW1 in work [10].

Here we consider more complicated cases – light passing through a perforated screen or an inhomogeneous film, or scattering on a planar crystal [11]. In these cases one momentum eigenstate in the input produces a superposition of an infinite number of different ES in the output. This type of EW was dubbed in [10] as EW2. The EW2 may include two distinct sets – one on the transmission side, and the other – on the incidence side. Both sets are formed in the near field (NF) and have common basic properties, so we may study only one of them. The diffraction experiments with light passing through a screen are usually focused on the far field (FF), so the EW2 remain overlooked.

A special variety of EW is the outer part of photon states propagating along a linear (or planar) dielectric waveguide. Some waveguides on microscopic level (atomic chains or planes) were considered in [12, 13]. Mathematically, the corresponding states are autonomous from any input, and their exponentially decreasing tails are side extensions of the guided wave. They are external parts of solution of a homogeneous equation for a waveguide, whereas EW1, 2 are special solutions of inhomogeneous equation with an external source (in particular, the EW2 is the NF solution of the Lippman-Schwinger equation [14, 15] for scattering on a screen considered as a macroscopic target). These distinctions warrant special name – EW3 – for the lateral tails of the guided waves[1].

The different types of EW are not always sharply defined, due to variability of experimental setups. For instance, EW1 can also be considered as a limiting case of EW3 for a homogeneous planar dielectric waveguide of thickness $b \to \infty$, when we can focus only on one side of the waveguide and regard the wave incident on this side as an external source. In the opposite limit $b \to 0$ we obtain the *surface waves* [16–18]. Another example (considered in Sec.3) is the evanescence transfer from the photon to a probing particle, which is accompanied by conversion of EW2 into a state similar to EW1 but without TIR.

Below we consider mostly EW2.

Let a plane wave with frequency $\omega_0$ be incident on a screen (e.g., grating) along the $z$-axis (Fig. 1). There emerges an infinite set of EW2 with different propagation numbers $k_x$, moving along the screen with different phase velocities $|u(k_x)| < c$. In an idealized model [10] – a grating with $N \to \infty$ infinitely narrow slits separated by a distance $d$ – the set $k_x$ is discrete, with eigenvalues $k_x \to k_x^{(m)}$ given by

$$k_x^{(m)} = mk_d, \quad |m| = 0, 1, 2, ..., \quad k_d \equiv \frac{2\pi}{d} \qquad (1.1)$$

---

[1]We do not include here possible EW with imaginary propagation number *along* a conductive waveguide. These could be called EW4, and are just the cavity modes below the threshold frequency.

The evanescence in this case is observed for sufficiently high $m$, such that

$$|m|k_d > k_0 \equiv \omega_0/c, \quad \text{or} \quad |m| > m_c \equiv \left[\frac{k_0}{k_d}\right] = \left[\frac{d}{\lambda_0}\right], \qquad (1.2)$$

where square-bracketed $[X]$ stands for the integer part of $X$. In all other systems the set of $k_x$ is continuous, and evanescence is associated with all $k_x$ satisfying the condition

$$|k_x| > k_0 \equiv \omega_0/c \qquad (1.3)$$

For such $k_x$ the $z$-component of the wave vector $\tilde{\mathbf{k}} = k_x\hat{\mathbf{x}} + \tilde{k}_z\hat{\mathbf{z}}$ is imaginary:

$$\tilde{k}_z = \pm i\chi_z, \quad \chi_z \equiv \sqrt{k_x^2 - k_0^2} \qquad (1.4)$$

Henceforth, the capped symbols will denote unit vectors, and the tilde above a symbol will indicate the possibility for it to be complex. If at least one component of a vector $\mathbf{A}$ can be complex or just imaginary, then the whole vector will also be written as $\tilde{\mathbf{A}}$. The "+" sign in (1.4) is taken for the transmission side, and the "$-$" sign for the incidence side. In either case, the corresponding phase velocities along the screen are

$$u(k_x) = \frac{\omega_0}{k_x}, \quad |u(k_x)| < c \qquad (1.5)$$

Their magnitudes are all less than $c$ by virtue of (1.3); and by the same token, the respective wavelengths

$$\lambda(k_x) = \frac{2\pi|u(k_x)|}{\omega_0} = \frac{2\pi}{|k_x|} < \lambda_0 \equiv \frac{2\pi}{k_0} \qquad (1.6)$$

are less than the input wavelength $\lambda_0$. In view of (1.5) we can call the EW "crawling waves" (CW) as opposed to regular running waves (RW). The same abbreviation CW can also mean "compressed waves" in view of (1.6). One can as well read the CW as "clinging waves" (they "cling" to the guiding surface where their amplitude is maximal). And in some cases, e.g., for EW2 on a conducting screen, the CW may be read as "cheating waves" because, while each one of them separately clings to the screen, they conspire to interfere destructively with departing waves for the whole observed set to give, say, the zero net field on the surface. Such richness of meaning warrants using the abbreviation CW (at least in English!) on the same footing as EW, so that farther in the text we will use both interchangeably.

In the limit $k_0 \to 0$ ($\lambda_0 \to \infty$), we will for any $k_x \neq 0$ have a CW with $u(k_x) \to 0$, so that even a one-way CW "freezes" (comes to a full stop). A "frozen wave" also obtains at finite $k_0$ in the limit $|k_x| \to \infty$.

For the case considered here (the input being a plane wave with $\mathbf{k}_0 = \omega_0/c\ \hat{\mathbf{z}}$), the output state above the screen (Fig. 2) can be written as

$$\Psi(\mathbf{r}, t) = 2\left\{\underbrace{\left(\int_0^{k_0} \mathcal{F}(k_x) e^{ik_z z} \cos k_x x\, dk_x\right)}_{\text{RW}} + \underbrace{\left(\int_{k_0}^{\infty} \mathcal{F}(k_x) e^{-\chi_z z} \cos k_x x\, dk_x\right)}_{\text{CW}}\right\} e^{-i\omega_0 t} \quad (1.7)$$

The first integral ($|k_x| \leq k_0$) is taken over the set of crossed RW receding upwards from the screen, and the second integral ($|k_x| > k_0$) is the superposition of all CW. We assumed that the aperture function of the screen peaks at its center and is symmetric about it, so the amplitude of a state $e^{i\mathbf{k}\mathbf{r}} = e^{i(k_x x + \tilde{k}_z z)}$ is an even function, $\mathcal{F}(k_x) = \mathcal{F}(-k_x)$, forming a system of standing waves along the screen. At the same time, the superposition (1.7) must describe a non-zero flux along the screen. This is similar to decomposition of a plane wave into spherical harmonics: each of the latter is a spherical standing wave, and yet the whole superposition describes a running wave. Since in our case the flux must head away from the screen's center on either side, we may have separate analytical expressions for $x < 0$ and $x > 0$.

A similar equation will describe the output on the incidence side ($z \leq -\delta$). It will differ from (1.7) by the sign of $\tilde{k}_z$ and, generally, by amplitudes $\mathcal{F}(k_x)$, so the whole output, when written in Cartesian coordinates, will be described by different analytical expression in each quadrant of plane $(x, z)$. A single analytical expression for the whole state *far from the screen* can be written in spherical coordinates. It describes scattering of the input wave by the screen (Fig. 3):

$$\Psi(\mathbf{r}, t) \sim \left[e^{ik_0 z} + f(\theta, \varphi)\frac{e^{ik_0 r}}{r}\right] e^{-i\omega t}, \quad r \gg D \quad (1.8)$$

Here $D$ is the maximal size of the screen, and $f(\theta, \varphi)$ is the scattering amplitude depending on polar and azimuthal angles $\theta$, $\varphi$.

Getting back to (1.7), one can, in principle, observe each term in the first integral in an appropriate measurement because each such term describes the corresponding RW receding from the screen, and asymptotically all such waves separate from each other. But there is no spatial separation between different CW. The analysis in [10] showed that separate modes of EW2 cannot be singled out in a NF-measurement, so the set of CW-s from the second sum in (1.7) can be observed only as an entire superposition. Such conclusion follows from the requirement that an accurate momentum measurement necessary to collapse this sum to a single momentum eigenstate must be performed on a free particle, whereas a EW2-photon is not free.

After having slipped off the screen's edge, all CW-s moving, say, to the left, merge into a single RW with wavelength $\lambda_0$ (some features of conversion from CW to RW were described qualitatively in [19]). This opens a possibility of their "retrospect" observation at least in the middle field (MF), and (arguably) in the FF. The converted RWs diverge from their "sources" (thin layers of EW "atmosphere" around the screen). Layers corresponding to different $|k_x| > k_0$ have different effective thickness $z_d \approx 1/\chi_z$ and the corresponding waves may be deflected into slightly different but overlapping angular regions of the opposite semi-space, so the respective plane components may also separate in the FF. But their amplitudes will generally be much less than for the elements of the first integral in (1.7). The same happens with the set moving to the right.

There is also a possibility of an indirect measurement of EW2-eigenstates by observing a probing particle after it had selectively absorbed the EW-photon in a certain eigenstate. Such observation can be made in the NF, or in the FF after the particle has slipped off the edge of the screen. The following sections present a more detailed analysis of some of these effects.

## 2. EW2 detection using a probing beam

Important experiments on scanning the profile (the *z*-dependence) of an EW1 were described in [1–3]. In [2], the EW1-photons were scattered from a small dielectric (polystyrene) sphere used as a probing particle near the interface. The scattered intensity as a function of *z* allows one to determine $z_d$.

Here we have EW2 instead of EW1, and we want to see if the whole superposition EW2 can be collapsed to a single eigenstate with some definite $k_x = k'_x$. We could do it by observing a probing particle selectively interacting with one of the superposed states. A NF experiment [3] using probing particles moving parallel to the interface in EW1 can in principle be used for probing EW2 as well. A reliable detection of EW photons is based on their absorption by the beam particles. An EW photon has some tachyonic properties [3] due to one imaginary component of its momentum. Therefore it can be absorbed (or emitted) even by a free electron [20]. Measuring the output electron could give us information about the state of absorbed photon.

A probing particle in the input state can be described as a Gaussian wave packet

$$\Phi(x,z) = \Phi_0 e^{-\frac{1}{2}\sigma^2(z-z_0)^2} e^{iK_0 x}, \qquad \sigma \equiv \left(\sqrt{2}\,\Delta z\right)^{-1} \qquad (2.1)$$

Here $\Delta z$ is the initial spread (standard deviation) along *z*; $z_0$ is the packet's distance from the plane $z = 0$, and $K_0$ is propagation number. As mentioned, electrons can absorb quanta of EW2; but they also interact with the screen through the image forces [21], and emit Smith-Purcell radiation [22], which complicates the experiment.

The probability of photon absorption in definite $k_x$-state is proportional to

$$\mathcal{P}(k_x) \sim \left| \mathcal{F}(k_x) \int_0^\infty e^{-\frac{1}{2}\sigma^2(z-z_0)^2} e^{-\chi_z z} dz \right|^2 \qquad (2.2)$$

(normalizing factor $|\Phi_0|^2$ dropped). For sufficiently high $|k_x|$ the shape of the packet (2.1) beyond the corresponding distance $z_d$ becomes immaterial, so we can set $\sigma = 0$ and approximate (2.2) by

$$\mathcal{P}(k_x) \underset{|k_x| \gg k_0}{\longrightarrow} \frac{|\mathcal{F}(k_x)|^2}{\chi_z^2} = \frac{|\mathcal{F}(k_x)|^2}{k_x^2 - k_0^2} \approx \frac{|\mathcal{F}(k_x)|^2}{k_x^2} \qquad (2.3)$$

The $\mathcal{P}(k_x) \to 0$ at high $k_x$. This cannot be remedied by preparing sharply localized beam with $\Delta z \approx z_d$. First, it will bring all the electrons close to the surface, with the above-mentioned consequences. Second, the corresponding state will undergo rapid quantum-mechanical (QM) spread along the $z$-direction. For packet (2.1) such spread is described by (see, e.g., [23])

$$\Delta z(t) = \Delta z \sqrt{1 + \frac{\hbar^2 t^2}{4\mu^2 (\Delta z)^4}} \quad, \qquad (2.4)$$

where $\mu$ is the rest mass of the electron. For a photon with $\lambda_0 = 0.5\,\mu m$ in an EW2-eigenstate with relatively small $|k_x|$ satisfying (1.3), say, $|k_x| = 5k_0$, we have from (1.4) $z_d \simeq (\chi_z)^{-1} \approx 0.2\lambda_0$. Let the probing electron's velocity be $v = 10^{-3} c$. In order to match the ES mode $|k_x| = 5k_0$, we prepare such electron with the initial $z$-indeterminacy $\Delta z \simeq z_d \approx 0.2\lambda_0$. For the 0.1-m distance between the electron source and the closest edge of the screen, it will take $t \approx 3 \times 10^{-7}\,s$ to reach the screen. Putting these data into (2.4) gives $\Delta z(t) = 1.5 \times 10^3 \Delta z$. Already at the start of probing, the $z$-expansion of the packet will exceed $\Delta z \simeq z_d$ of the EW photon by more than 3 orders of magnitude. This takes us back to situation described by Eq. (2.3).

A better option may be using sufficiently heavy ions or neutral atoms, with appropriate velocities and transition frequencies. Their main advantage is the change of their inner state and thereby of their rest mass after absorption/emission of an EW-photon, which gives more flexibility in the experimental options as described in [3].

Now we turn from these technical issues to fundamental principles which determine the basic features of EW2 interactions with environment.

### 3. Evanescence transfer to a probing particle

One of the fundamental features in the EW-photon absorption by a probing particle is transfer of the imaginary component of photon's momentum to the particle. *This throws the particle itself into an EW-state*, regardless of whether the particle is elementary or not. We assume the probing particle to be an electron, without loss of generality of the basic results.

The EW- photon absorption by an electron is determined by conservation laws

$$\Omega = \Omega_0 + \omega_0; \quad \tilde{\mathbf{K}} = \mathbf{K}_0 + \tilde{\mathbf{k}} = (K_0 + k_x)\hat{\mathbf{x}} + \tilde{k}_z \hat{\mathbf{z}}, \qquad (3.1)$$

.

where the capital symbols $\Omega$, $\tilde{\mathbf{K}}$ indicate the electron's energy and momentum (in units of $\hbar$). We assume the initial condition $\mathbf{K}_0 = K_0 \hat{\mathbf{x}}$ (an incident electron state is a monochromatic plane wave along the *x*-direction). If $|k_x| > k_0$, the electron acquires an imaginary $\tilde{K}_z$ originating from absorbed $\tilde{k}_z$. This means that the *evanescence is transferred to the electron in the process of absorption*. The problem of the NF-detection of an EW2-photon is converted to a problem of detection of the EW-electron.

We can exclude $\tilde{k}_z$ from (3.1) by using, in addition to (1.4), also the dispersion equations for the initial and final state of the electron

$$\Omega^2 - c^2 \tilde{\mathbf{K}}^2 = \Omega_0^2 - c^2 K_0^2 = \Omega_\mu^2, \quad \Omega_\mu \equiv \frac{\mu c^2}{\hbar} \qquad (3.2)$$

This gives equation for the eigenvalue $k_x$ of a photon-state that can be absorbed

$$\left(\frac{\Omega_0}{c} + k_0\right)^2 - (K_0 + k_x)^2 - (k_0^2 - k_x^2) = \frac{\Omega_\mu^2}{c^2} \qquad (3.3)$$

It follows

$$k_x = \frac{\Omega_0 \omega_0}{c^2 K_0} = k_0 \sqrt{1 + \frac{K_\mu^2}{K_0^2}} \quad (|k_x| > k_0), \quad K_\mu \equiv \frac{\mu c}{\hbar}, \qquad (3.4)$$

where the square root is taken with the same sign as $K_0$. Result (3.4) automatically satisfies the condition for evanescence of the absorbed photon. This is natural, since only an ES- photon can be absorbed by the electron. The absorbed $\tilde{k}_z$ is

$$\tilde{k}_z = \sqrt{k_0^2 - k_x^2} = k_0 \sqrt{-\frac{K_\mu^2}{K_0^2}} = i k_0 \frac{K_\mu}{K_0} \qquad (3.5)$$

Therefore the electron's momentum right after the photon absorption will be

$$\tilde{\mathbf{K}} = \left(K_0 + k_0 \sqrt{1 + \frac{K_\mu^2}{K_0^2}}\right)\hat{\mathbf{x}} + i k_0 \frac{K_\mu}{K_0} \hat{\mathbf{z}} \qquad (3.6)$$

It is immediately seen that the electron itself is now in an EW-state. But it is a single EW-eigenstate rather than their superposition. The photon absorption by the electron is accompanied by transition EW2 → EW1 (but the latter without TIR!).

Eq-s (3.3, 4) select either $k_x$ for a given $K_0$, or $K_0$ for a given $k_x$. For the idealized model [10] with discrete set (1.1) of $k_x$, the values of $K_0$ determined by (3.3) will also form a discrete set. The selection will be possible only under condition

$$k_x^{(m)} = k_0\sqrt{1+\frac{K_\mu^2}{K_0^2}} = mk_d, \quad \text{or} \quad K_0 \to K_0^{(m)} = \frac{K_\mu}{\sqrt{m^2\left(\frac{k_d}{k_0}\right)^2 - 1}} \quad (3.7)$$

for a given integer $|m| > m_c$. If this condition is not satisfied, no EW2-eigenstates will be suitable for absorption. But in all realistic situations the spectrum of $k_x$ is continuous, so requirements (3.7) do not apply, and for any electron with $\mathbf{K}_0 = K_0\hat{\mathbf{x}}$ and input photon with $\mathbf{k}_0 = k_0\hat{\mathbf{z}}$ there is *a CW-eigenstate* in (1.7) which can be absorbed by this electron. There is only one exception, when $K_0 = 0$ (the electron is initially at rest, $\Omega_0 \to \Omega_\mu$), and $\omega_0 \to 0$. Then Eq-s (3.3, 4) will be satisfied by any finite $k_x$ – any CW mode in (1.7) can be absorbed. The electron does not gain any energy after absorption: even though it acquires a non-zero momentum $K_x = k_x$ along the x-direction, the corresponding energy increase is exactly balanced by its loss due to the absorbed imaginary $k_z$-momentum. Loss equals the gain at $\omega_0 \to 0$. This gives rise to an interesting possibility of spontaneous formation of EW2 photon states without any input photons, and their transfer to a stationary electron.

Except for the case $K_0 = 0$, Eq-s (3.4, 6) rule out a possibility of using both branches (folds) of the electron dispersion curve (surface) in the $K_x$, $\Omega$-plane (Fig. 4a, b) for possible solutions. Gaining energy $\omega_0$ means shifting up the surface. Remaining on the same branch means acquiring $k_x$ of the same sign as $K_0$; jumping to the opposite branch requires large change of momentum $|\Delta K_x| = |k_x| > |2K_0|$. But this brings in a large negative energy due to accompanying $k_z$-component (Fig. 4b), which would make the output energy less than $\Omega_0 + \omega_0$. So absorption of an EW- photon can accelerate the electron but not turn it back. A moving electron can only absorb an EW photon crawling in the same direction. The most important features of this process are:
1) The absorption is selective: an electron with given $K_0$ "picks up" a photon's ES state with a *definite* $k_x$ (and respective $\tilde{k}_z$), adding them to its initial momentum;
2) The electron itself becomes evanescent after this;
3) Collapse of superposition (1.7) to a *single* EW-*eigenstate* for the electron produces a state similar to an EW1.

Properties 2) and 3) appear to contradict the requirement that before the absorption, the *whole set* (1.7) is needed for the EW2-photon to satisfy the boundary condition on the screen. But there is no inconsistency here because interaction with the surface and the initial and boundary conditions for the electron differ from those of the photon; the electron gets into contact with the screen by coming from aside instead of passing through it. This makes it far less sensitive to the slits which are crucial for the state of EW2-photon.

As mentioned above, the electron-EW1-state emerging according to 3) differs from EW1 as defined in the beginning of Sec. 1 by being autonomous from TIR which is just absent here.

In any case, the electron will eventually slide off the screen's edge and recede to the FF domain. But particle's momentum in FF is entirely real – the electron regains its "regular" status described by de Broglie's wave. It gets rid of its $\tilde{K}_z$-component and changes momentum from (3.6) to one with the magnitude

$$K = \sqrt{c^{-2}(\Omega_0 + \omega_0)^2 - K_\mu^2} \qquad (3.8)$$

corresponding to the free electron. Expressed in terms of $K_0$, the succession of events *for the electron* (input, selective absorption of the EW2-photon, and exiting into the FF) can be described as

$$\underbrace{K_0 \hat{\mathbf{x}}}_{\text{Input}} \rightarrow \underbrace{\left(K_0 + k_0\sqrt{1 + \frac{K_\mu^2}{K_0^2}}\right)\hat{\mathbf{x}} + ik_0 \frac{K_\mu}{K_0}\hat{\mathbf{z}}}_{\text{Intermediate state, NF}} \rightarrow \underbrace{\left(\sqrt{K_0^2 + 2k_0\sqrt{K_\mu^2 + K_0^2} + k_0^2}\right)\hat{\mathbf{n}}}_{\text{Output, FF}} \qquad (3.9)$$

Here $\hat{\mathbf{n}}$ is a unit vector representing the average direction of the output electron, which depends on geometry of the edge and generally tilts towards the opposite semi-space. Both actions (absorption of an EW2-photon in the NF and the following conversion to a free-electron state in the FF) are consecutive stages of one process. Note that $K$ is $k_x$-independent. Therefore its measured value in the output only indicates, in retrospect, the fact of absorption of the EW2-photon. Information about specific absorbed eigenstate $|k_x\rangle$ is lost, although it may still be contained in the *direction* $\hat{\mathbf{n}}$.

Another option is to skip the probing and just allow the EW2-photon itself to slide off the screen, with the same effect – conversion to the regular photon. Since in this case the intermediate state is the superposition (1.7), it is more convenient to express all stages in terms of the eigenstates rather than eigenvalues

$$\underbrace{|k_0\hat{\mathbf{z}}\rangle}_{\text{Input}} \rightarrow \underbrace{\int \mathcal{F}(k_x)|\tilde{\mathbf{k}}(k_x)\rangle dk_x}_{\text{Intermediate state, NF}} \rightarrow \underbrace{|k_0\hat{\mathbf{n}}\rangle}_{\text{An output, FF}} \qquad (3.10)$$

The final stage here is a state to which the scattered wave chooses to collapse in an appropriate momentum measurement, and the unit vector $\hat{\mathbf{n}}$ has the same meaning as in (3.9), albeit with a wider range of possible directions. Technically, observing the EW2 photon in this process would look as detecting photon scattered by an angle $\theta > 90^\circ$ (e.g., found in the lower semi-space in an experiment shown in Fig. 2, 3, when the bottom face of the screen is made absorbing to eliminate the light scattering from it).

Both options may open a new venue in the study of EW2-states. It includes calculating the angular distribution of the output wave in the FF for a given screen.

Summarizing this part, we can say that for each input state (the momenta $\mathbf{k}_0$, $\mathbf{K}_0$, respectively, of the incident photon and probing electron), Eq-s (3.4, 5) select the value $k_x$ from the spectrum of the intermediate EW2-photon that can be absorbed with probability (2.2), and Eq. (3.6) determines the resulting complex-valued momentum $\tilde{\mathbf{K}}$ of the intermediate electron. Such $\tilde{\mathbf{K}}$ cannot be directly measured in the NF, and converts into the real-valued vector $K\hat{\mathbf{n}}$ in the FF. But since its norm is $k_x$-independent, the possibility of experimental confirmation of the selected $k_x$ remains an open question. A FF-measurement possibly could give us *some* information about $k_x$ from orientation of $\hat{\mathbf{n}}$.

In all cases the question arises – where does the imaginary component of the electron/photon's momentum go after their sliding off the screen? By the same token, one could ask – where does the imaginary component of an EW-photon momentum come from, to begin with? And where does the change (3.9) (or $\tilde{\mathbf{k}} \to k_0\hat{\mathbf{n}}$ in (3.10)) go to?

The electron/photon momentum may change during their interaction not only with each other, but also with the screen. This involves the screen into the process of energy-momentum exchange.

### 4. Photon-screen interaction

Here we will focus on the energy-momentum exchange between the photon and the screen, so **K** will stand for the *screen*'s momentum. Our first question is: will such exchange allow one to observe the EW-photon momentum eigenstates by measuring **K** before and after the interaction?

Since the set of $k_x$ is continuous, we must measure **K** with arbitrarily high precision. Such measurement brings the screen close to de Broglie's state $|K_x\rangle$, with position indeterminacy $\Delta x \to \infty$ for its center. It is equivalent to "smearing out" its transparency profile over *x*, that is, making the screen effectively homogeneous and accordingly producing no ES! And vice versa, retaining screen's "profile" requires $\Delta x$ to be less than profile's characteristic half-width $\delta_x$. This leads to $\Delta K_x \geq \hbar/2\delta_x$ already in the input state, even if the average $\overline{K_x} = 0$. For a screen with sharply defined slits this demands its input state with $\Delta K_x \to \infty$. The same holds for the $K_z$-component: all the output waves must know the exact *z*-coordinate of the top and bottom face of the screen in order to satisfy the corresponding boundary conditions. The $K_x$-indeterminacy blocks the possibility to extract information about individual $|k_x\rangle$-eigenstates of the EW2-photon by momentum measurement of the screen. This confirms the result [10] about inseparability of different EW2-eigenstates in the NF from a somewhat different perspective.

Next, we will focus on quantitative description of the photon-screen interaction.

For all practical purposes, it would be safe to consider the screen's mass as infinite (as we did in the previous sections), thus neglecting vanishingly small changes of screen's, and accordingly, the photon's energy. This would greatly simplify the equations, practically without any loss of accuracy. Nevertheless, now we will go beyond this approximation. This will make the analysis more complete and will give a deeper insight into the whole phenomenon. In particular, it will expose the connection between ES and the Gamow states (GS) in the system.

Suppose that the screen is loosely connected with the Lab (imagine an experiment in a non-rotating space station with engines off). Then we can consider it as an autonomous object of mass $M$. We can measure **K**, e.g., by measuring frequency of monochromatic light reflected from the screen.

In all known physical situations, $Mc^2 \gg \varepsilon_0 \equiv \hbar\omega_0$, but once we take $M$ to be finite, the screen after the interaction acquires, together with momentum, a small kinetic energy $E$. Accordingly, the photon loses equal amount of energy, so in its ES its energy $\tilde{\varepsilon} \neq \varepsilon_0$.

At this point, we must emphasize an important distinction between vector space $V$ for observable vectors like momentum, and the Hilbert space $\mathcal{H}$ for quantum states. The norm of a state vector $|\Psi\rangle = \sum c_n |n\rangle$ in $\mathcal{H}$ is given by $\langle\Psi|\Psi\rangle = \sum c_n^* c_n$. In contrast, according to dispersion equation (1.4), the norm of a physical vector $\tilde{\mathbf{k}} = \sum \tilde{k}_j \hat{\mathbf{x}}_j$ in $V$ is given by $\tilde{k}^2 = \sum \tilde{k}_j^2$, *not* $\langle\tilde{k}|\tilde{k}\rangle = \sum \tilde{k}_j^* \tilde{k}_j$, *even when some of the components $\tilde{k}_j$ are complex*, which is the case in the ES. Then the "magnitude" $\tilde{k} = \sqrt{\sum \tilde{k}_j^2}$ is not necessarily a real (let alone definite positive) number (the word "magnitude" here is in quotation marks to emphasize that the corresponding quantity can be complex due to complex $\tilde{k}_z$). This has dramatic implications when we calculate $E$ for the screen with momentum $\tilde{\mathbf{K}}$.

Consider a photon input state $|\mathbf{k}_0\rangle$ incident on the screen as shown in Fig. 1. The photon gets entangled with screen on the pre-measurement stage. We assume that the QM average of screen's momentum in its initial state $|S_0\rangle$ is zero. The input state of the whole system is $|\mathbf{k}_0\rangle|S_0\rangle$, and pre-measurement output is an entangled superposition:

$$|\mathbf{k}_0\rangle|S_0\rangle \Rightarrow \int \mathcal{F}(k_x)|\tilde{\mathbf{k}}\rangle|\tilde{\mathbf{K}}\rangle dk_x \qquad (4.1)$$

Here $\tilde{\mathbf{k}} = k_x \hat{\mathbf{x}} + \tilde{k}_z \hat{\mathbf{z}}$ and $\tilde{\mathbf{K}} = K_x \hat{\mathbf{x}} + \tilde{K}_z \hat{\mathbf{z}}$ are momentum eigenvalues for the photon and screen, respectively, and the kets are their eigenstates. The integral on the right contains both – RW-s ($|k_x| \leq k_c$) and CW-s ($|k_x| > k_c$) (the critical value $k_c$ will be calculated later). Expression (4.1) generalizes (1.7) by including screen's states. In position representation

$$\langle\mathbf{r}|\tilde{\mathbf{k}}\rangle \equiv \psi_{\tilde{\mathbf{k}}}(\mathbf{r}) = e^{i\tilde{\mathbf{k}}\mathbf{r}} = e^{i(k_x x + \tilde{k}_z z)}, \quad \langle\mathbf{R}|\tilde{\mathbf{K}}\rangle \equiv S_{\tilde{\mathbf{K}}}(\mathbf{R}) = e^{i\tilde{\mathbf{K}}\mathbf{R}} = e^{i(K_x X + \tilde{K}_z Z)} \qquad (4.2)$$

Here $\mathbf{R}$ is position vector of screen's center of mass, and $\psi_{\mathbf{k}}(\mathbf{r})$, $S_{\mathbf{K}}(\mathbf{R})$ are the wave functions of the photon and screen, respectively (normalizing factors dropped). In these notations (4.1) can be written as

$$\psi_{\mathbf{k}_0}(\mathbf{r})S_0(\mathbf{R}) \Rightarrow \int \mathcal{F}(k_x)\psi_{k_x}(\mathbf{r}) S_{k_x}(\mathbf{R})dk_x = \int \mathcal{F}(k_x)e^{i\tilde{\mathbf{k}}\mathbf{r}}e^{i\tilde{\mathbf{K}}\mathbf{R}}dk_x , \qquad (4.3)$$

where

$$\tilde{\mathbf{K}} \equiv \mathbf{k}_0 - \tilde{\mathbf{k}} = -k_x\hat{\mathbf{x}} + (k_0 - \tilde{k}_z)\hat{\mathbf{z}} \qquad (4.4)$$

Here $\tilde{k}_z$ and thereby $\tilde{\mathbf{k}}$ and $\tilde{\mathbf{K}}$ are (yet unknown) functions of given $k_x$. Thus, a $|\tilde{\mathbf{k}}\rangle$-photon state originates by losing the initial momentum $\mathbf{k}_0$ and acquiring in exchange a new momentum $\tilde{\mathbf{k}}$. The screen's state $|\tilde{\mathbf{K}}\rangle$ originates by acquiring momentum $\mathbf{k}_0$ from the input photon and giving $\tilde{\mathbf{k}}$ to the output photon, so the resulting momentum of the screen is $\tilde{\mathbf{K}} = \mathbf{k}_0 - \tilde{\mathbf{k}}$. The initial net momentum $\mathbf{k}_0$ of the whole system conserves in each individual term of superposition (4.1).

In view of the above-mentioned fact that $\tilde{\varepsilon} \neq \varepsilon_0$, we also have

$$\tilde{k} = \sqrt{k_x^2 + \tilde{k}_z^2} \neq k_0 \qquad (4.5)$$

for any $k_x \neq 0$. The "magnitude" of screen's momentum in eigenstate $|\tilde{\mathbf{K}}\rangle$ is

$$\tilde{K} \equiv \sqrt{K_x^2 + \tilde{K}_z^2} = \sqrt{k_x^2 + (k_0 - \tilde{k}_z)^2} = \sqrt{k_0\left(k_0 - 2\sqrt{\tilde{k}^2 - k_x^2}\right) + \tilde{k}^2} \qquad (4.6)$$

Since $\tilde{k}_z$ can be imaginary and, as we will see shortly, generally complex in an ES, the corresponding $\tilde{K}_z$ will also be complex, with nonzero real and imaginary parts. This answers the question raised in the end of the previous section. The imaginary part of $\tilde{k}_z$ is opposed by the equal part of $\tilde{K}_z$. Both are created simultaneously in the process of interaction when $|k_x| > k_c$. Similarly, the non-zero $k_x$ originates from interaction with the screen which recoils along $x$ with momentum $K_x = -k_x$, thus playing the role of the probing particle. In the final stage when the EW-photon/electron slides off the screen, and the $x$-component of its momentum undergoes change of the type (3.10) from CW to RW, the screen undergoes symmetric change, recoiling in the opposite direction.

For a non-dissipative process assumed here, the conservation of energy requires that

$$\tilde{E} = \varepsilon_0 - \tilde{\varepsilon} \equiv \hbar c(k_0 - \tilde{k}) \equiv \delta\tilde{\varepsilon} \qquad (4.7)$$

(where $\delta\tilde{\varepsilon}$ is the *change* of photon's energy) or

$$\sqrt{M^2c^2 + \hbar^2\left(k_0^2 - 2k_0\sqrt{\tilde{k}^2 - k_x^2} + \tilde{k}^2\right)} - Mc = \hbar\left(k_0 - \tilde{k}\right) \qquad (4.8)$$

Here we have an important distinction from system (3.3), (3.4). In the former, with the screen immovable, we choose the initial momenta $\mathbf{k}_0$, $\mathbf{K}_0$ for the photon and probing particle, respectively, and find the corresponding $k_x$ at which that photon can be absorbed by the particle. In (4.8), with screen itself serving as the probing particle with the zero initial momentum, we need to find the *energy* $\tilde{\varepsilon} = c\tilde{k}$ *of the output photon* for given $k_0$ and *chosen* $k_x$. Thus, for each $k_x$, the corresponding $\tilde{k}$ must be found as the solution of Eq. (4.8)[1]. After some algebra we obtain

$$\tilde{k} = \frac{\mathcal{K}_M + k_0 \pm \sqrt{k_0^2 - \left(1 + 2\dfrac{k_0}{\mathcal{K}_M}\right)k_x^2}}{\mathcal{K}_M + 2k_0} k_0, \qquad \mathcal{K}_M \equiv \frac{Mc}{\hbar} \qquad (4.9)$$

In the simplest case $k_x = 0$ (the photon passes through the screen without deflection) there is no 4-momentum exchange with the screen, so there must be $\tilde{k} = k_0$. Therefore out of two solutions (4.9), only one with the "+" sign is physically acceptable.

For $\delta\tilde{\varepsilon}$ (and accordingly for energy $\tilde{E}$ acquired by the screen), we obtain

$$\delta\tilde{\varepsilon} = \tilde{E} = \hbar c \, \frac{k_0 - \sqrt{k_0^2 - \left(1 + 2\dfrac{k_0}{\mathcal{K}_M}\right)k_x^2}}{\mathcal{K}_M + 2k_0} k_0 \qquad (4.10)$$

In the limit $M \to \infty$, Eq-s (4.9, 10) for any finite $|k_x|$ reduce to familiar $\tilde{k} = k_0$, $\tilde{E} = 0$. Finally, the same calculation gives

$$\tilde{k}_z = \sqrt{\tilde{k}^2 - k_x^2} = \frac{k_0^2 + (\mathcal{K}_M + k_0)\sqrt{k_0^2 - \left(1 + 2\dfrac{k_0}{\mathcal{K}_M}\right)k_x^2}}{\mathcal{K}_M + 2k_0} \qquad (4.11)$$

We see that the condition for evanescence changes from $|k_x| > k_0$ to

$$|k_x| > k_c \equiv \eta k_0, \quad \eta \equiv \sqrt{\frac{\mathcal{K}_M}{\mathcal{K}_M + 2k_0}} \qquad (4.12)$$

---

[1] One could ask why do we use the relativistic expression for screen's energy? Non-relativistic approximation would still be extremely accurate in the considered case, and the initial equation would look simpler than (4.8). But its solution would be much more complicated! We have here a rare case when more general (and accordingly more accurate) equation is solved much simpler than its approximation.

This defines the critical value $k_c$ separating RW from CW (the same is seen already from (4.9)). And more important, when this condition is satisfied, the $\tilde{k}_z$ has apart from imaginary also a small but non-zero real part, so we can write (4.11) in the form:

$$\tilde{k}_z = \frac{k_0^2}{\mathcal{K}_M + 2k_0} + i\chi_z, \quad \chi_z \equiv \eta\left(1 + \frac{k_0}{\mathcal{K}_M}\right)\sqrt{k_x^2 - \eta^2 k_0^2} \qquad (4.13)$$

According to (4.4), this leads to complex $\tilde{K}_z$:

$$\tilde{K}_z = k_0\left(1 - \frac{k_0}{\mathcal{K}_M + 2k_0}\right) - i\chi_z \qquad (4.14)$$

As a result, the energies of both – photon and screen – in any ES also become complex:

$$\tilde{\varepsilon} = \hbar c \tilde{k} = \varepsilon + i\gamma, \quad \tilde{E} = E - i\Gamma \qquad (4.15)$$

Here

$$\varepsilon \equiv \frac{\mathcal{K}_M + k_0}{\mathcal{K}_M + 2k_0}\varepsilon_0, \quad E = \frac{k_0}{\mathcal{K}_M + 2k_0}\varepsilon_0, \qquad (4.16)$$

and

$$\gamma = \Gamma \equiv \frac{\chi_z}{\mathcal{K}_M + k_0}\varepsilon_0 \qquad (4.17)$$

with $\chi_z$ given by (4.13). The real parts of the output photon and screen's energy are both $k_x$-independent. Their imaginary part $\Gamma$ increases with $|k_x|$ and may dominate over $\varepsilon$ and especially over $E$. In all practical cases we have $\hbar\omega_0 \ll Mc^2$, so $E \ll \hbar\omega_0$. As to $\Gamma$, it may be arbitrarily large. For sufficiently high $|k_x| \gg k_c$, Eq. (4.17) gives

$$\Gamma = \frac{\hbar^2}{M}k_0\chi_z \approx \frac{\hbar^2}{M}k_0|k_x|, \qquad (4.18)$$

so $\Gamma$ is unbounded.

### 5. Connection between ES and the Gamow states

States with complex energy are known as the Gamow states (GS). They are usually formed within a region enclosed by a spherical potential barrier, and decay by tunneling through it. The corresponding states are observed in radioactive nuclei, resonance scattering, or resonance particles in high energy physics [24-26]. Here we obtained the possibility of a similar state for a totally different system which, under the above assumptions, can be considered as purely classical and essentially free. For a free object, complex energy describes the spread of its initially localized wave packet (or its time-reverse), and in our case it happens in the classical limit. The spread of the packet is accompanied by probability decrease in its central part and increase in the peripheral parts. In this respect, it is similar to decay of the initially localized state. Under condition (4.18) the average life-time of such state is

$$\tau = \frac{\hbar}{\Gamma} \to \frac{M}{\hbar k_0 |k_x|} \qquad (5.1)$$

In realistic situations with macroscopic screen and reasonable values of $|k_x|$, the corresponding values of $\tau$ are very large. For instance, consider a typical case

$$M = 0.1 \text{ kg}, \quad k_0 = 7.5 \times 10^6 \text{ m}^{-1} \text{ ("red" photon in the input), and } |k_x| = 10 k_0 \qquad (5.2)$$

Suppose that in a momentum measurement of the screen, the intermediate state (entangled superposition (4.3)) collapsed to a state defined by (5.2). Then (5.1) gives $\tau \approx 1.8 \times 10^{18} \, s \approx 6 \times 10^{10} \, Y$. This is 3 times longer than the age of observable universe, so the screen can still be described classically (that is, $\Gamma$ is negligible). But one could argue that according to (5.1), the $\tau$ becomes arbitrarily small at sufficiently large $|k_x|$, and the screen would then "disappear" from its initial region almost instantly. Such a behavior would conflict with the concept of classical object.

There are two counter-arguments to it. First, we must take momentum into the picture. The complex energy (4.15) and momentum (4.13), (4.14) are two sides of the same coin. Taking account of both is especially important in the 1-D case [23] when we have evanescence only along the $z$-direction. According to (4.14), the real part of $\tilde{K}_z$ is positive, that is, the screen gets a kick up. At the same time, its imaginary part and therefore its evanescence is opposite in sign to that of the photon. Its probability amplitude exponentially decreases *down* the $z$-direction, and this is observed when screen's center of mass shifts up. The "disappearance" in this case is due, apart from the QM spread as such, also to the shifting of the wave packet as a whole in the $+z$ direction, so that the next moment the less intensive part of its tail passes by the observer. And its fast rate does not necessarily mean high speed of the packet: in view of its extremely small width (it is actually a sharp "spike"), it can "disappear" nearly instantly from its initial location even at slow motion. This effect contributes to the rapid disappearance of the screen at high $|k_x|$. As an example, consider (4.14) at $\mathcal{K}_M \gg k_0$, and $|k_x| \gg k_c$. This gives $\tilde{K}_z \approx k_0 - i\chi_z \approx k_0 - i|k_x|$. The upward velocity of the screen is $v_z \approx \hbar k_0 / M$, and the packet width is $d_z \approx 1/\chi_z \approx 1/|k_x|$. Then the traverse time will be

$$\tau(k_x) = \frac{d_z}{v_z} \approx \frac{M}{\hbar k_0 |k_x|} \qquad (5.3)$$

This is identical to (5.1)! So the short lifetime under given conditions is mostly due to motion of the asymmetric "spike" (in our case the EW-tail of the *moving* screen) rather than to its spread. The whole effect is similar to evolution of a narrow wave packet as described in [23].

Second, we must also take into account the corresponding probabilities. The probability for state (4.3) to collapse to $k_x$-th mode rapidly falls off with increase of $|k_x|$. In the simplest case of a screen with one slit of width $a$, for the photon with the lowest TE-mode within the slit [21] we have

$$|\mathcal{F}(k_x)|^2 = 4\pi a \frac{\cos^2 \frac{1}{2} a k_x}{\left(\pi^2 - a^2 k_x^2\right)^2} \tag{5.4}$$

The probability to find *any* $k_x$ higher than some value $k_x'$ greatly exceeding $k_a \equiv \pi/a$ is

$$\widetilde{\mathcal{P}}(k_x) = 2\int_{k_x'}^{\infty} |\mathcal{F}(k_x)|^2 \, dk_x \underset{k_x' \gg k_a}{\approx} \frac{8}{3} \frac{\pi}{(ak_x')^3} \tag{5.5}$$

Comparison with (5.1) shows that probability to observe a small lifetime of the packet drops much faster than time itself. This is a typical feature of the classical limit of QM. There is a non-zero probability for a tennis ball to tunnel through a 1-meter thick concrete wall. The question is only how small this non-zero value is. In other words, how long is the expectation time for such probability to actualize. Evaluation from known expression for transmission probability through the corresponding potential barrier will give the time exceeding the age of the universe. This is again similar to result given by Eq. (5.1).

On the other hand, one can always find the parameters of the studied system for which the range of $k_x$ associated with sufficiently low lifetimes will be represented with sufficiently high probability. An obvious example follows from the same Eq-s (5.3-5) for sufficiently small $a$. But even in such cases, the result (5.3) shows that the short lifetime must be attributed mostly to the motion of the screen as a whole rather than the spread of its wave packet.

Now the similar questions arise about the photon: how to interpret 1) the real part of $\tilde{k}_z$ and 2) the imaginary part of $\tilde{\varepsilon}$, whose sign is, according to (4.14), opposite to that of the screen.

Property 1) can be crudely visualized as a secondary effect: the screen gets a "kick up" (momentum $k_0 \hat{\mathbf{z}}$) from the absorbed input photon, and as a result "pushes up" the output photon emerging on its opposite side. This gives rise to a very small real part of $\tilde{k}_z$ in (4.13). Another way to put it: the $\tilde{k}_z$ in an EW-state is purely imaginary in the reference frame $S_M$ attached to the screen. This frame itself starts moving together with screen relative to the Lab in the output state. But once the EW state of the photon has been formed, the screen, albeit stationary in $S_M$, acquires the opposite imaginary component $-i\chi_z$ of momentum. Performing now the Lorentz transformation of the photon 4-momentum from $S_M$ back to the Lab frame, we obtain the transformed $\tilde{k}_z$ with non-zero real part and $\tilde{\varepsilon}$ with non-zero imaginary part. The same transformation for the screen 4-momentum gives complex $\tilde{E}$ and $\tilde{K}_z$.

Property 2) means the *increase* of any local probability density $\left|\langle \mathbf{r}|\mathbf{k}\rangle_{\text{Ph}}\right|^2$ with time for the output photon in the $k_x$-th mode for $|k_x| > k_c$. This is consistent with the initial assumptions: the screen's location was initially set in semi-space $z \leq 0$, and an output photon in the chosen set of CW2 exists in $z \geq 0$. Accordingly, the photon probability density measured by a stationary observer may increase due to contribution from more intense parts of its exponential tail approaching from beneath.

Thus, with screen's mobility included in the theory, the system (screen + output photon) in its $k_x$-th mode is described by

$$\psi_{k_x}(\mathbf{r}, t) S_{k_x}(\mathbf{R}, t) \sim e^{i\left(\tilde{\mathbf{k}}\mathbf{r} - \frac{\tilde{\varepsilon}}{\hbar}t\right)} e^{i\left(\tilde{\mathbf{K}}\mathbf{R} - \frac{\tilde{E}}{\hbar}t\right)}, \tag{5.6}$$

and the whole superposition will be

$$\Psi(\mathbf{r}, \mathbf{R}, t) = \int_{k_x} \mathcal{F}(k_x) \psi_{k_x}(\mathbf{r}, t) S_{k_x}(\mathbf{R}, t) dk_x \sim \int_{k_x} \mathcal{F}(k_x) e^{i\left(\tilde{\mathbf{k}}\mathbf{r} - \tilde{\varepsilon}t\right)} e^{i\left(\tilde{\mathbf{K}}\mathbf{R} - \tilde{E}t\right)} dk_x \tag{5.7}$$

But in view of (4.7), the total energy in each product (5.6) remains $\varepsilon_0$. Each term of superposition (5.7) describes the corresponding state of a system of two objects (screen + photon) with fixed net energy $\varepsilon_0 = \hbar\omega_0$, only differently split between the objects in each eigenstate. Therefore the whole superposition (5.7) reduces to

$$\Psi(\mathbf{r}, \mathbf{R}, t) \sim \left\{ \int_{k_x} \mathcal{F}(k_x) e^{i\tilde{\mathbf{k}}\mathbf{r}} e^{i\tilde{\mathbf{K}}\mathbf{R}} dk_x \right\} e^{-i\omega_0 t}, \tag{5.8}$$

which allows one to bring it to the form similar to (1.7).

The situation described by Eq-s (5.7, 8) is reminiscent of some known systems with their energy eigenstates being each a superposition of non-stationary states. A good example is an $NH_3$-molecule – each of its 2 stationary states with indefinite location of the N-atom is a simple superposition of 2 non-stationary states with definite locations [23]. Here we have a more general case of a complex system (photon + screen) in an *entangled* superposition of an infinite number of non-stationary states. And, in contrast with $NH_3$, in which two basic non-stationary states have the same decay rates, here all the non-stationary elements of superposition (5.7) have different decay rates, while the system is still stationary in its whole. But the most fundamental distinction from $NH_3$-type states is that EW2 eigenstates are not separable from each other in the NF.

The whole effect is specific only to EW2 due to existence of arbitrarily high $\chi_z$. It is absent in EW1, where $\chi_z$ has a relatively small fixed value $\chi_z = k_0\sqrt{n^2\sin^2\theta - 1}$ for the given refraction index $n$ and $\theta$ exceeding critical angle of incidence $\theta_c = \text{Arcsin}(1/n)$.

## 6. Conclusions

**6.1** Formation of EW2 is a complicated phenomenon involving superposition of momentum eigenstates with complex vector eigenvalues. But it does not allow to single out directly a definite EW2-eigenstate. Only the whole superposition of them can be observed in the NF. The underlying cause for such non-separability is the photon interaction with inhomogeneous screen, which requires a set of eigenstates to satisfy the boundary condition on the surface. Such a problem does not exist in EW1 where the interface between the two media is uniform and produces a single output eigenstate for each input eigenstate. Experimentally, the EW1 state manifests itself already in the NF, e.g., in the Goos-Hänchen effect.

**6.2** Interaction between the EW2-photon and environment involves a complicated process of 4-momentum exchange. One of its characteristics is evanescence transfer to a probing particle which does not interact with the screen and thus can select and inherit a single EW2-photon eigenstate. This allows one to study the EW2-eigenstates indirectly by selectively transferring them to a probing particle with an appropriate momentum. The selection is in this case totally deterministic – the transferred eigenstate is uniquely determined by the input states of the photon and probing particle.

**6.3** The ES and the GS are two sides of the same coin since **K** is a spatial part of a 4-vector $(\Omega/c, \mathbf{K})$. If at least one of the **K**-components is complex-valued ($\mathbf{K} \to \tilde{\mathbf{K}}$), then $\Omega$ will be also complex-valued ($\Omega \to \tilde{\Omega}$) in some RF-s, and vice-versa. Both observables are coupled by Lorentz transformations. Formation of EW2 on a movable screen considered in Sec. 4 shows physical details of the ES-GS connection.

**6.4** As a by-product of the whole discussion, we see the possibility of indirect study of EW2 by observing their imprint in the NF or FF. Such imprint may be produced by an EW2-photon sliding off an edge of the screen and receding to a region free of RW. A simple experiment could be placing detectors in the regions inaccessible for the transmitted or reflected RW appearing in the given setup. Referring to Fig. 2, 3, it may be the semi-space below the absorbing bottom face of the screen. In this case there are no reflected waves, and detection of a photon below the screen's plane would be a clear sign of EW2 converted into the secondary RW in the MF and/or FF.

(*Note*: Some aspects of this work were presented in an abstract [27] in a Poster-Session of 2014 DAMOP meeting)

## Acknowledgements
I am grateful to Henk Arnoldus for valuable comments.

**Figures**

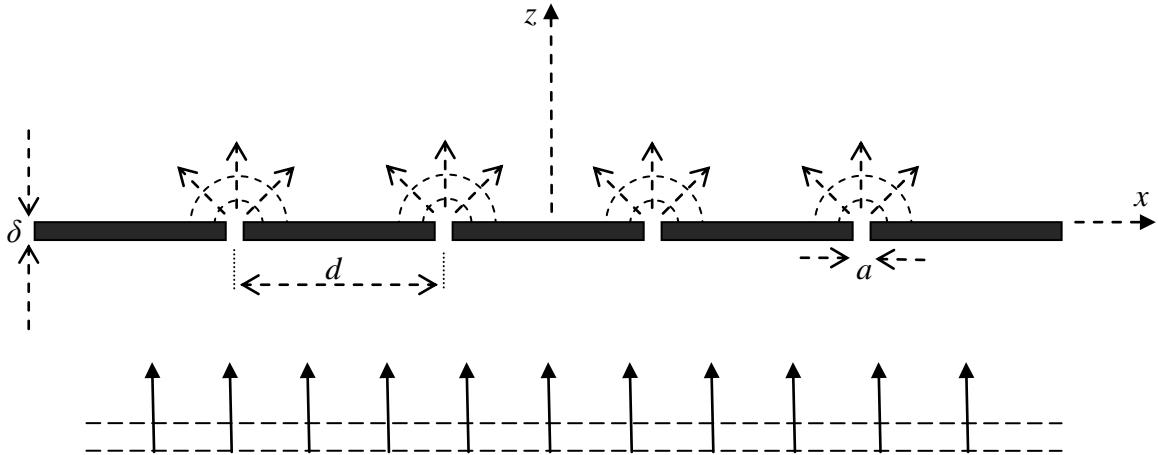

**Fig. 1**
Schematic of the multiple interference experiment with a screen as diffraction grating

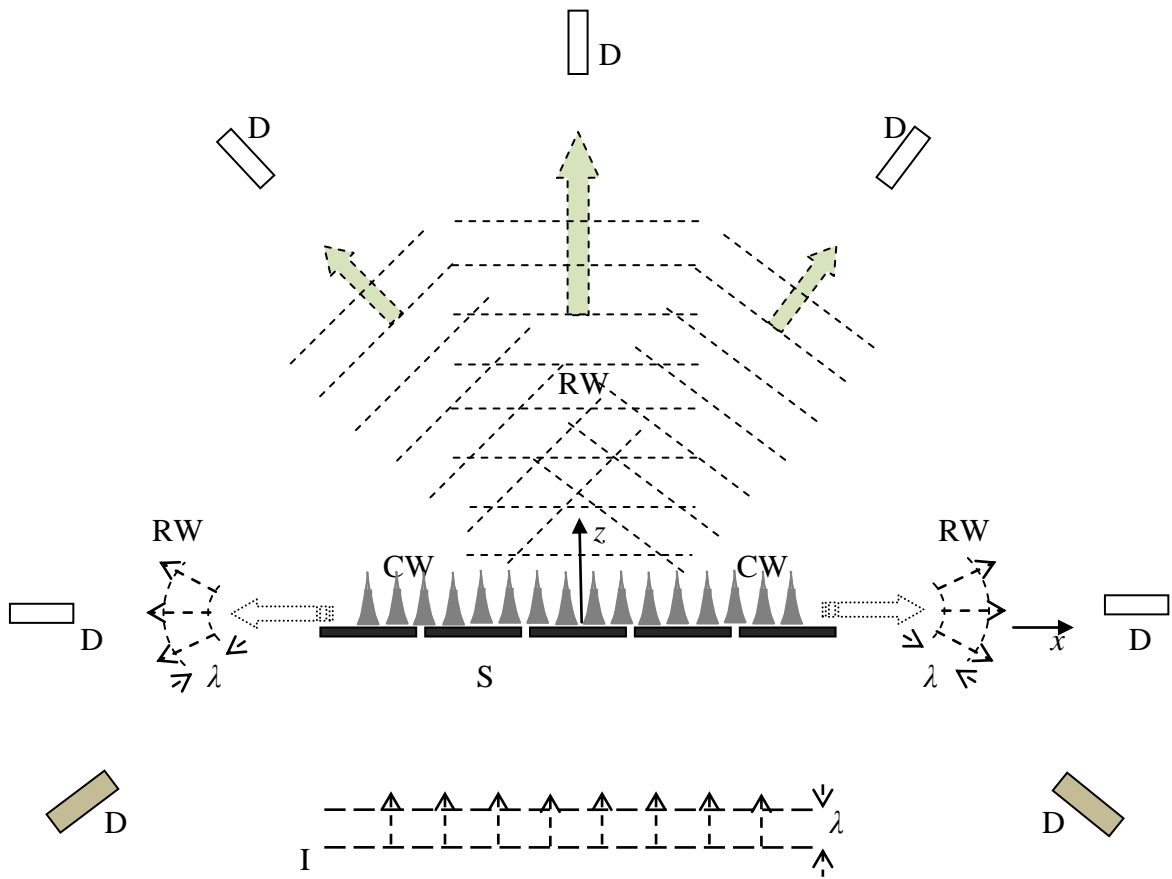

**Fig. 2**
The RW and EW (CW) in the output from the incident plane monochromatic wave I.
S – screen;  D - detectors.
 The CW are shown only for one definite value of $k_x$. A similar system may form on the incidence side. On the left and right side of the screen, each set of CW converts into one RW with the same wavelength $\lambda_0$ as in the input wave. These secondary RW can be considered as the imprints of the initial CW into the FF. The shaded detectors in the lower semi-space can record such imprints, especially in the absence of RW reflected from the incidence side, which would mask the converted waves.

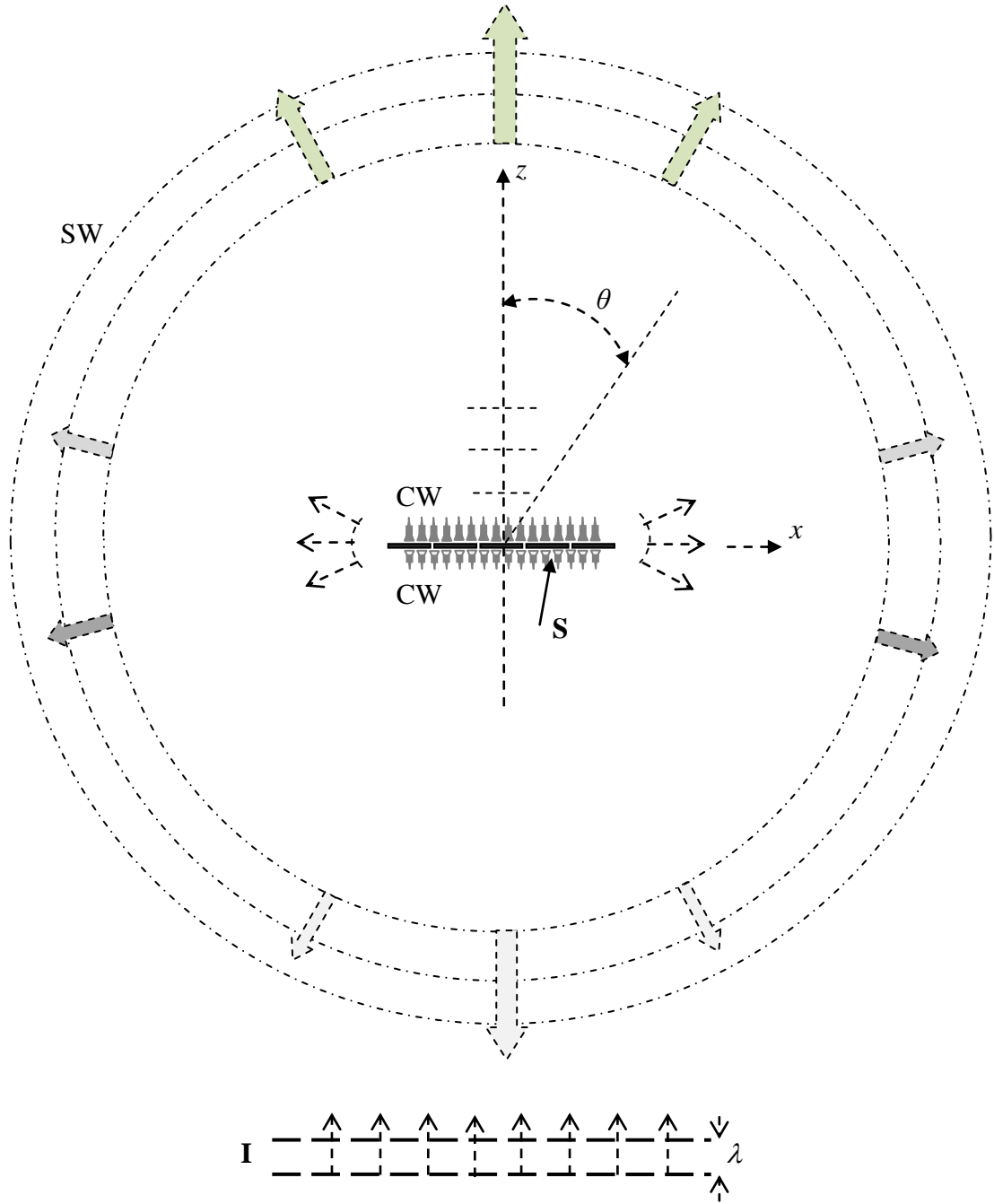

**Fig. 3**
Schematic of a monochromatic scattering process (not to scale). **I** – input (incident wave). S – screen (scatterer); CW – crawling waves (for some definite $k_x$) in the NF right above and below the screen; SW – scattered wave (final output in the FF) that can be described as diverging spherical wave with direction-dependent amplitude $f(\theta, \varphi)$. The $\varphi$-dependence is not shown

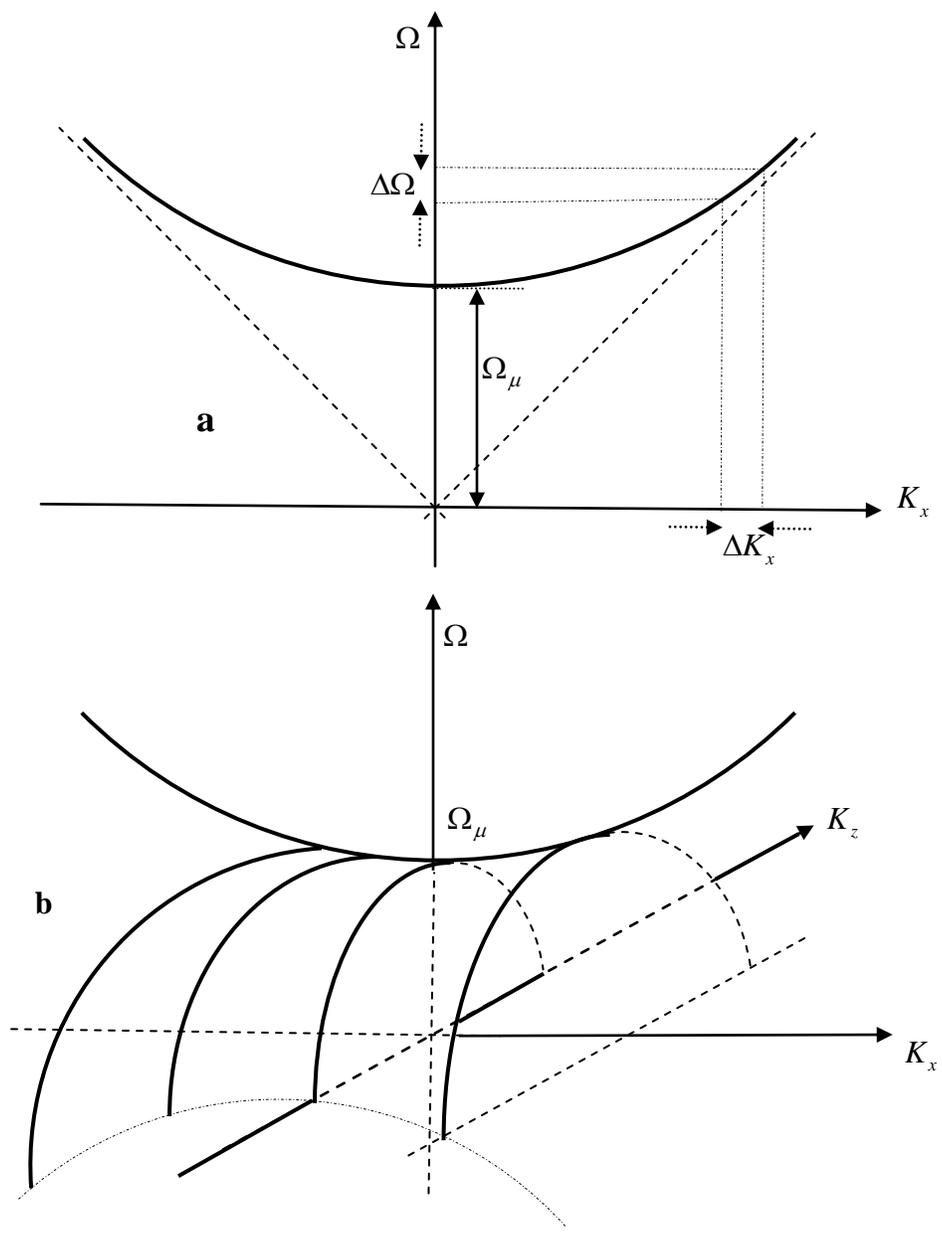

**Fig. 4**
(a) Graph representing the conservation laws (3.1) and dispersion equation (3.2) for a free electron, and a possible change of its energy and momentum after absorption of an EW-photon.
(b) 2-D saddle-like surface of negative curvature in momentum space representing the dispersion equation for an ES-electron.